\documentclass[12pt,preprint]{aastex}
\usepackage{graphicx}
\newcommand{\be}{\begin{equation}}
\newcommand{\ee}{\end{equation}}
\newcommand{\bea}{\begin{eqnarray}}
\newcommand{\eea}{\end{eqnarray}}
\begin{document}
\title{MHD Turbulence as a Foreground for CMB Studies}
\author{Jungyeon Cho, \& A. Lazarian}
\affil{Dept. of Astronomy, University of Wisconsin,
    475 N. Charter St., Madison, WI53706; cho, lazarian@astro.wisc.edu}

\begin{abstract}

Measurements of intensity and polarization of diffuse Galactic synchrotron
emission as well as starlight polarization reveal power law
spectra of fluctuations. We show that these fluctuations
can arise from magnetohydrodynamic (MHD) turbulence in the Galactic
disk and halo. To do so we take into account the
converging geometry of lines of sight for the observations when
the observer is within the turbulent volume. 
Assuming that 
the intensity of turbulence changes along the line
of sight, we get a reasonable fit to the observed synchrotron
data. As for the spectra of polarized starlight we get 
a good fit to the observations taking into account the
fact that the observational sample is biased toward
nearby stars.

\end{abstract}


\section{Introduction}

Attempts to determine the statistics of intensity and polarization
fluctuations of cosmic microwave background (CMB) renewed
interest to the fluctuations of Galactic foreground radiation
(see Tegmark et al.~2000).
Spectra of intensity of synchrotron emission and synchrotron polarization
(see papers in de Oliveira-Costa \& Tegmark 1999)
as well as starlight polarization 
(Fosalba et al. 2002; henceforth FLPT)
have been measured. Those measurements revealed a range of power-laws,
the origin of which has not been addressed as far as we know. In
Tegmark et al.~(2000) there is an allusion that the spectra may be
relevant to Kolmogorov turbulence, but the issue of how those different spectra
may arise has not been addressed.

Interstellar medium is turbulent and Kolmogorov-type spectra
were reported on the scales from several AU to several kpc
(see Armstrong et al. 1995; Lazarian \& Pogosyan 2000; Stanimirovic
\& Lazarian 2001). Therefore it is natural to think of the 
turbulence as the origin of the fluctuations of the diffuse foreground
radiation. Interstellar medium is magnetized with magnetic
field making turbulence anisotropic. 
It may be argued that although the spectrum of
MHD turbulence exhibits scale-dependent anisotropy if
studied in the system of reference defined by the local
magnetic field (Goldreich \& Sridhar 1995; Lithwick \&
Goldreich 2001; Cho \& Lazarian 2002),
in the observer's system of reference the spectrum will
show only moderate scale-independent anisotropy. Thus
from the observer's point of view Kolmogorov description of
interstellar turbulence is acceptable in spite of the fact that
magnetic field is dynamically important and even dominant (see discussion in 
Lazarian \& Pogosyan 2000; Cho, Lazarian \& Vishniac 2002).  
  
It is customary for
CMB studies to expand
the foreground intensity  over spherical harmonics $Y_{lm}$,
        $I(\theta, \phi)= \sum_{l,m}a_{lm}Y_{lm}(\theta,\phi)$, 
and write the spectrum in terms of  
        $C_l\equiv \sum_{m=-l}^{m=l} |a_{lm}|^2/(2l+1)$.
The measurements indicate that 
angular power spectra ($C_l$) of the Galactic emission
follows power law ($C_l\propto l^{-\alpha}$)
(see FLPT and references in \S3).

This paper tests whether the measured spectra are compatible
with the  theoretical
predictions of spatial statistics that is expected in the presence of
MHD turbulence.
Analytical studies in this direction were
done in Lazarian (1992, 1994, 1995) and here we supplement them
with numerical simulations of the synthetic spectra.

\section{Spectra of Fluctuations}

\subsection{$C_l$ for small angle limit}
\begin{figure*}[!t]
\plotone{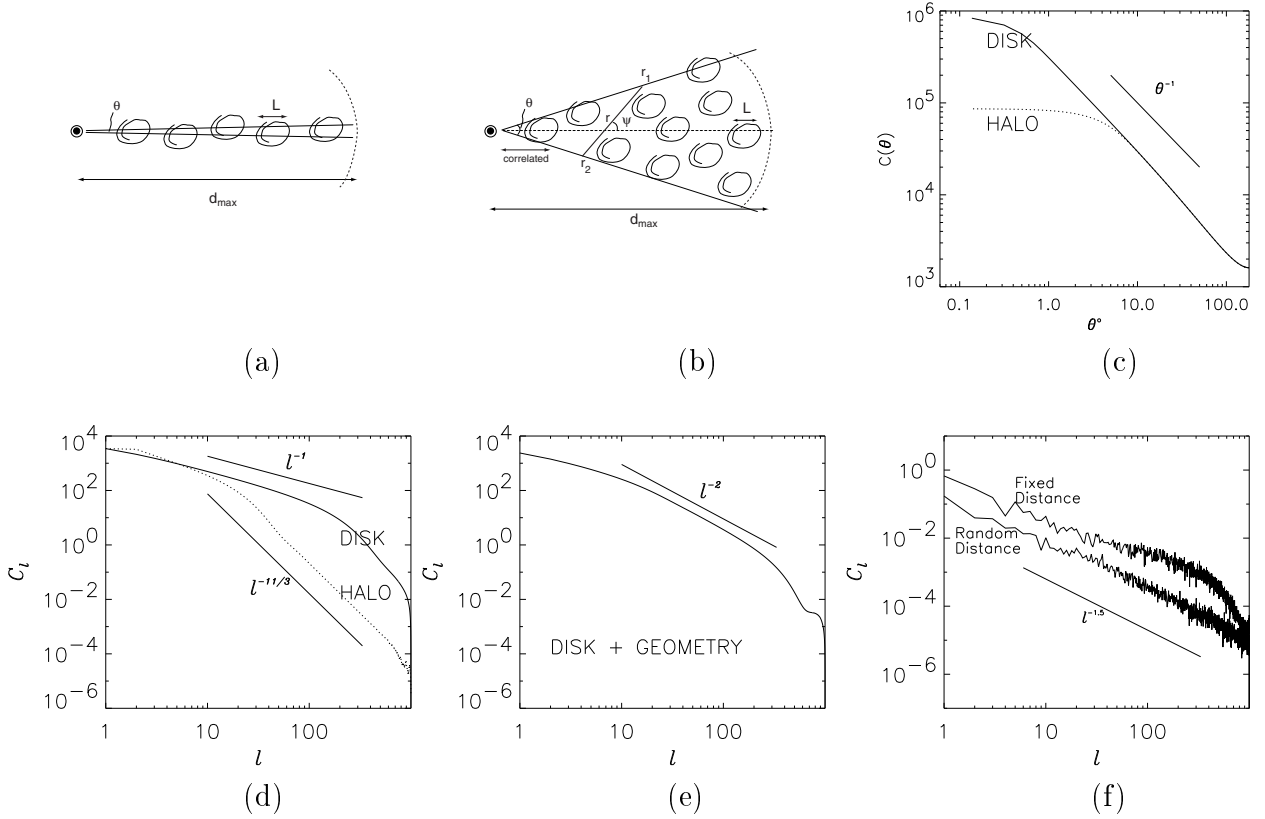}
\caption{
          (a) Small $\theta$ limit ($\theta <L/d_{max}$). The fluctuations
 along the entire length of the lines of sight are correlated.
          (b) Large $\theta$ limit ($\theta >L/d_{max}$). 
              Only points close to the observer 
 are correlated.
          (c) Angular correlation function for disk (solid line) and
          for halo (dotted line).
          When $\theta$ is small, $K(\theta)\propto constant-\theta^{5/3}$.
          When $\theta$ is large, $K(\theta)\propto \theta^{-1}$.
          (d) Spectra for disk ($d_{max}$ = 10 kpc; turbulence size L = 100 pc)
              and for halo ($d_{max}$ = 1 kpc; turbulence size L = 100 pc).
          (e) Angular spectrum for Galactic disk. We considered geometry of 
              Galactic disk.
          (f) Angular spectra of degree of polarization.
          Fixed Distance means all star are at the same distance.
          Random Distance means random sampling of stars according to equation
          (\ref{eq_13}).
          Zero point of the y-axis is arbitrary for all graphs.
         }
   \label{fig_all}
\end{figure*}

In this section we show that, when the angle between the lines of sight
is small (i.e.~$\theta < L/d_{max}$), the
angular spectrum $C_l$ has the same slope as the 3-dimensional
energy spectrum of turbulence.
Here $L$ is the typical size of the largest energy containing eddies,
which is sometime called as outer scale of turbulence or energy injection 
scale,
and
$d_{max}$ is the distance to the farthest eddies (see Fig.~\ref{fig_all}a).

To illustrate the problem consider the observations with lines of sight being
parallel. The observed intensity is the intensity summed
along the {\it line of sight}, $r_z$.
\begin{eqnarray}
 {I}_{2D}(r_x,r_y) & \equiv & \int d{r_z}~i_{3D}({\bf r})    \label{2d3d} \\
 & = & \int d{r_z}\ \int dk_x dk_y dk_z\ \hat{i}_{3D}({\bf k})\
  e^{i{\bf k}\cdot{\bf r}}.
\end{eqnarray}
Rearranging the order of summation and using 
$\int d{r_z}\ e^{ik_zr_z}=\delta(k_z)$, we get
\begin{equation}
  {I}_{2D}(r_x,r_y) = 
  \int d{k_x} dk_y ~\hat{i}_{3D}(k_x,k_y,0)   ~e^{ik_xr_x+ik_yr_y},
\end{equation}
which means Fourier component of $ {I}_{2D}(r_x,r_y)$
is $\hat{i}_{3D}(k_x,k_y,0)$. 

When the angular size of the observed region 
($\Delta \theta \times \Delta \theta$ in radian)
on the sky is small, 
$C_l$ is approximately the `energy' spectrum of ${I}_{2D}(r_x,r_y)$
(Bond \& Efstathiou 1987; Hobson \& Majueijo 1996; Seljak 1997), which means
$C_l \sim |\hat{i}_{3D}(k_x,k_y,0)|^2$ with $l\sim k (\pi/\Delta \theta)$
and $k=(k_x^2+k_y^2)^{1/2}$.
The analysis of the geometry of crossing lines
of sight is more involved, but for power-law statistics it follows from
 Lazarian \& Shutenkov (1990) that if $|\hat{i}_{3D}|^2\propto k^{-m}$, 
then the `energy' spectrum
of  ${I}_{2D}(r_x,r_y)$ is also $k^{-m}$.  Therefore, we have 
\begin{equation}
   C_l \propto  |\hat{i}_{3D}(k_x,k_y,0)|^2 \propto l^{-m}.
\end{equation}
in the small $\theta$ limit\footnote{
    In some cases, we infer $C_l$ from the observation of 
    the correlation function ${\cal K}(r)$
    in real space (or $K(\theta)$ on the sky).
    When the three-dimensional spectrum of turbulence 
    follows a power law ($\sim k^{-m}$),
    ${\cal K}(r) \propto {\cal K}_0-r^{m-2}$, where ${\cal K}_0\sim L^{m-2}$
    is a constant.
    However, when the slope of the turbulence spectrum is 
    steeper than $k^{-4}$,
    this relation breaks down and 
    ${\cal K}(r) \propto {\cal K}_0-r^{2}$ regardless
    of the turbulence slope.
    When we infer $C_l$ from this ${\cal K}(r)$, we obtain $C_l\propto l^{-4}$ 
    regardless of the true slope, when the slope is steeper than $-4$.
}.

For Kolmogorov turbulence ($|\hat{i}_{3D}|^2\propto k^{-11/3}$),
we expect
\begin{equation}
C_l \propto l^{-11/3}, \mbox{~~~if $\theta<L/d_{max}$.}
 \label{eq_5}
\end{equation}
Note that $l\sim \pi/\theta$.

\subsection{$C_l$ for large angle limit}

{}Following Lazarian \& Shutenkov (1990), we can show that
the correlation function for $ \theta > L/d_{max}$,
\begin{eqnarray}
    K(\theta) & = & \int \int dr_1 dr_2 ~{\cal K}( |{\bf r}_1-{\bf r}_2| ), \nonumber
\\
              & = & \frac{1}{\sin{\theta}}
                    \int_0^{\infty}dr ~r {\cal K}(r) \int_0^{\pi/2}d\psi 
\sim \frac{const}{\theta},
\end{eqnarray}
where we change variables: $(r_1,r_2)\rightarrow (r,\psi)$ is clear from
Fig.~1b; we accounted for the Jacobian of which is $r/\sin{\theta}$.
We can understand $1/\theta$ behavior qualitatively from Fig.~1b.
When the angle is large, points along  of the lines-of-sight near the observer
are still correlated. These points extend from the observer
over the distance $\propto 1/\sin{(\theta/2)}$.

In the limit of $\theta\ll 1$ 
we get the angular power spectrum $C_l$ 
using Fourier transform:
\begin{eqnarray}
   C_l & \sim &  \int \int  K(\theta) 
                 e^{-i{\bf l}\cdot {\bf \theta}} d\theta_x d\theta_y  
\nonumber \\
       & \sim &  \int d\theta ~\theta J_0(l\theta) K(\theta) \propto   l^{-1},
   \label{eq_7}
\end{eqnarray}
where $\theta=(\theta_x^2+\theta_y^2)^{1/2}$, $J_0$ is the Bessel function, and
we use $K(\theta)\propto \theta^{-1}$.

In summary, for Kolmogorov turbulence, we expect from equations (\ref{eq_5}) and
(\ref{eq_7}) that
\begin{equation}
 C_l \propto \left\{ \begin{array}{ll} 
                         l^{-11/3}     & \mbox{if $l>l_{cr}$} \\
                         l^{-1}        & \mbox{if $l<l_{cr}$.}
                      \end{array}
              \right.
     \label{eq_1_11_3}
\end{equation}
which means that the power index $\alpha$ of $C_l$ is\footnote{ 
      Note that point sources would result in $\alpha \sim 0$.} 
$-1 \leq \alpha \leq -11/3$. 
The critical number $l_{cr} ~(\sim 3d_{max}/L)$ 
depends on the size of the size of the large
turbulent eddies and on the direction of the observation. If in the
naive model we assume that turbulence is homogeneous along the lines
of sight and has $L\sim 100\ pc$ corresponding to a typical size of the
supernova remnant for disk with $d_{max}\sim 10$~kpc, we get $l_{cr}\sim
300$. For the synchrotron
halo, $d_{max}\sim 1$~kpc (see Smoot 1999) and we get $l_{cr}\sim
30$.

\section{Synchrotron Radiation}

Recent statistical studies of {\it total} synchrotron intensity
include Haslam all-sky map at 408 MHz that
shows that the power index $\alpha$ is in the 
range between
2.5 and 3 (Tegmark \& Efstathiou 1996; Bouchet, Gispert, \& Puget 1996).
Parkes southern Galactic plane survey at 2.4 GHz suggests shallower slope:
Giardino et al.~(2002) obtained $\alpha \sim 2.05$ after point source removal 
and Baccigalupi et al.~(2001) obtained $\alpha \sim -0.8$ to $-2$.
Using Rhodes/HartRAO data at 2326 MHz, Giardino et al.~(2001) obtained 
$\alpha\sim 2.43$ for all-sky  data and $\alpha\sim 2.92$ for
high Galactic latitude regions with $|b|>20^o$. The rough tendency
that follow from these data is that $\alpha$ which is  close to $2$
for the Galactic plane gets steeper (to $\sim -3$) for higher latitudes.
In other words $-2\leq \alpha \leq -3$ which differs from naive expectations
given by equation (\ref{eq_1_11_3}).

Can the difference be due to non-linear law of synchrotron emissivity?
For synchrotron radiation, emissivity at a point ${\bf r}$,
   $i({\bf r}) \propto n(e) |{B}_{\bot}|^{\gamma}$,
where $n(e)$ is the electron number density, $B_{\bot}$ is the component
of magnetic field perpendicular to the line of sight. The index
$\gamma$ is approximately $2$ for radio synchrotron frequencies
(see Smoot 1999).
If electrons
are uniformly distributed over the scales of magnetic field inhomogeneities,
the spectrum of intensity reflects the statistics of magnetic field.
For small amplitude perturbations
($\delta b/B\ll 1$; this is true for scales several times 
smaller than $L$ when we interpret $B$ as local mean magnetic field strength), 
if $\delta b$
has a power-law behavior, the statistics of intensity will have the 
same power-law behavior (see Deshpande et al.~2000). 
Therefore emissivity non-linearity does not account
for the difference between the observed spectra and the theoretical
expectations.

To address the problem, we perform simple numerical calculations
for galactic disk and halo.
We obtain $C_l$ using the relation
\begin{eqnarray}
 & &K(\cos \theta) = \frac{1}{4\pi}\sum_{l} (2l+1)C_l P_l(\cos{\theta}), \\
 & &C_l = \frac{1}{2} \int P_l(\cos{\theta}) K(\cos \theta)\ d(\cos\theta).   
                                                             \label{c_l_leg}
\end{eqnarray}
We use Gauss-Legendre quadrature integration (see Szapudi et al.~2001 for its
application to CMB) to obtain $C_l$.
We numerically calculate the angular correlation function from
\begin{equation}
    K(\cos{\theta})=\int dr_1 \int dr_2 ~{\cal K}(|{\bf r}_1-{\bf r}_2|),
   \label{eq_ctheta}
\end{equation}
where $|{\bf r}_1-{\bf r}_2|=r_1^2+r_2^2-2r_1r_2 \cos{\theta}$ and
assume
that 
the spatial correlation function  ${\cal K}(r)$ follow Kolmogorov statistics:
\begin{equation}
 {\cal K}(r) \propto \left\{ \begin{array}{ll} 
                              {\cal K}_0-r^{2/3} & \mbox{if $r<L$} \\
                              0          & \mbox{if $r>L$,}
                      \end{array}
              \right. \label{c_r_3D}
\end{equation}
where ${\cal K}_0\sim L^{2/3}$ is a constant.
Fig.~\ref{fig_all}c and  Fig.~\ref{fig_all}d 
illustrate the agreement of our calculations with the
theoretical expectations within the naive model of the disk and the
halo from the previous section.

To make the spectrum closer to  observations we need
to consider more realistic models.
First, synchrotron emission is stronger in spiral arms, and therefore
we have more 
synchrotron emission coming from the nearby regions.
Second, if synchrotron disk component is sufficiently
thin,
then lines of sight are not equivalent and effectively nearby
disk component contributes more. Indeed,
when we observe regions with low Galactic latitude,
the effective line-of-sight varies with Galactic latitude.

Suppose we observe a region with $b=10^o$.
Then emission from $d=10~kpc$ is substantially weaker than
that from $d=100~pc$, because the region with $d=10~kpc$ is $10\ kpc~\sin{10^o}
\sim 1.7~kpc$ above the Galactic plane and, therefore, has weak emission.
To incorporate this effect, we use the weighting function
$w(r)=100/max( 100, r \sin{10^o})$, which
gives more weight on closer distance.
The resulting angular power spectrum (Fig.~\ref{fig_all}e) 
shows a slope similar to -2.

For halo, the simple model predicts that $C_l \propto l^{-11/3}$, but
observations provide somewhat less steep spectra.
Is this discrepancy very significant?
The spectrum of magnetic field is expected to be
shallower than $k^{-11/3}$ in the vicinity of the energy injection
scale and at the vicinity of the magnetic equipartition scale. 
The observed spectrum also gets shallower if $d_{max}$ gets larger.
For instance 
Beuermann, Kanbach, \& Berkhuijsen (1985) reported the existence of
thick radio halo that extends to more than $\sim 3~kpc$ near the Sun.
Finally, filamentary structures and point sources
can make the spectrum shallower as well. Further research should establish
the true reason for the discrepancy.

\section{Galactic Starlight Polarization}
Polarized radiation from dust is an important
component of Galactic foreground that strongly interferes with
intended CMB polarization measurements (see Lazarian \& Prunet 2001). 
FLPT attempted to predict the spectrum of the polarized
foreground from dust and 
obtained $C_l \sim l^{-1.5}$ for starlight polarization.
This spectrum is different from those discussed in the previous sections.
To relate this spectrum to the underlying turbulence we should account
for the following facts: a) the observations are done for the disk component
of the Galaxy, b) the sampled stars are at different distances from
the observer with most of the stars close-by.

To deal with this problem we use numerical simulations again.
We first generate a three (i.e.~{\it x,y}, and {\it z}) 
components of magnetic field on a two-dimensional
plane ($4096 \times 4096$ grid points representing $10~kpc \times 10~kpc$), 
using the following Kolmogorov three-dimensional spectrum:
$E_{3D}(k)\propto k^{-11/3}$ if $k>k_0$,
where $k_0 \sim 1/100~pc$. Our results show that how we continuously
extend the spectrum  for $k<k_0$ does not change our results.

To simulate the actual distribution of stars within the sample
used in FLPT, we scatter our emission sources
using the following
probability distribution function:
\begin{equation}
   P(r) \propto e^{-r/1.5kpc}.    \label{eq_13}
\end{equation}
The starlight polarization is due to the difference in absorption
cross section of non-spherical grains aligned with their longer axes
perpendicular to magnetic field (see review by Lazarian 2000). In numerical
calculations we approximate the actual turbulent
magnetic field by a superposition
of the slabs with locally uniform magnetic field in each slap
and assume that the difference
in grain absorption parallel and perpendicular to magnetic field results
in the 10\% difference in the optical
depths $\tau_{\|}$ and $\tau_{\bot}$ for a slab. 
We calculate evolution of Stocks parameters of the starlight
within the slab and use the standard
transforms of Stocks parameters from one slab to another 
(Dolginov, Gnedin, \& Silantev 1996; see similar 
expressions in Martin 1972).

We show the result in Fig.~\ref{fig_all}f.
For comparison, we also calculate the degree of polarization assuming
all stars are at the same distance of $\sim10~kpc$.
The result shows that, for a mixture of nearby and faraway stars,
the slope steepens and gets very close to the observed one, i.e.
$-1.5$.

Our conclusions may be tested if stars at particular distance only are
correlated. If those stars are nearby, we would expect to have steeper
index and it will become shallow if only distant stars are chosen.
Choosing only nearby stars we expect to get a steeper index.
A systematic study of this change can provide an estimate
of the energy injection scale $L$.
{}From the point of view of foreground studies, we conclude that a
naive use of the 
polarization template produced with the random sample of stars may
be misleading.

\section{Summary}
In this paper we have addressed the origin of spatial
fluctuations of emissivity and polarization
of Galactic diffuse emission.
We have shown that MHD turbulence can qualitatively explain observed
properties of total synchrotron emission and starlight polarization.
The variety of measured spatial spectra of synchrotron emission
 can be explained by the
inhomogeneous distribution of emissivity, 
structure of Galactic disk and halo,
and/or various energy injection scales.
Similarly, MHD turbulence plus inhomogeneous distribution of stars 
can explain the reported scaling of starlight polarization
statistics.

Evidently more systematic studies are required.
Those studies will not only give insight into how to
separate CMB from foregrounds, but also
would provide valuable information on
the structure of interstellar medium and the sources/energy
injection scales of interstellar turbulence.

\acknowledgements 
We acknowledge the support of NSF Grant AST-0125544.

\end{document}